\documentclass[superscriptaddress,floatfix,amsmath,amssymb,aps,prx,reprint]{revtex4-2}
\usepackage{physics}
\usepackage{bbold}
\usepackage{natbib}    
\usepackage{color}         
\usepackage{graphicx}      
\usepackage{epsfig}          
\usepackage{bm}            
\usepackage{amsmath}
\usepackage{thumbpdf}
\usepackage[belowskip=-8pt,aboveskip=0pt]{caption}
\usepackage{subcaption}
\usepackage{hyperref} 
\usepackage{placeins}
\usepackage{xcolor}
\usepackage{multirow}
\usepackage[font=small,justification=justified,format=plain]{caption}
\usepackage[normalem]{ulem}
\usepackage{tikz}
\usetikzlibrary{calc}
\newcommand{\tikzmark}[1]{\tikz[overlay,remember picture] \node (#1) {};}
\newcommand{\DrawBox}[1][]{%
    \tikz[overlay,remember picture]{
    \draw[red,#1]
      ($(left)+(-0.2em,0.9em)$) rectangle
      ($(right)+(0.2em,-0.3em)$);}
}
\begin{document}

\title{Local Magnetoelectric Effects as a Predictor of Surface Magnetic Order}
\author{Sophie F. Weber}
\affiliation{Materials Theory, ETH Z\"{u}rich, Wolfgang-Pauli-Strasse 27, 8093 Z\"{u}rich, Switzerland}
\author{Andrea Urru}
\affiliation{Department of Physics and Astronomy,
Rutgers, the State University of New Jersey, Piscataway, New
Jersey 08854, USA}
\author{Nicola A. Spaldin}
\affiliation{Materials Theory, ETH Z\"{u}rich, Wolfgang-Pauli-Strasse 27, 8093 Z\"{u}rich, Switzerland}

\date{\today}
\begin{abstract}
We use symmetry analysis and density functional theory to show that changes in magnetic order at a surface with respect to magnetic order in the bulk can be generically determined by considering local magnetoelectric responses of the crystal. Specifically, analysis of the atomic-site magnetoelectric responses, or equivalently the corresponding local magnetic multipoles, can be used to predict all surface magnetic modifications arising purely from symmetry lowering via termination of the bulk magnetic order. This analysis applies even in materials with no bulk magnetoelectric response or surface magnetization. We then demonstrate our arguments for two example antiferromagnets, metallic $\mathrm{CuMnAs}$ and rock-salt $\mathrm{NiO}$. We find that the $(010)$ and $(1\bar{1}0)$ surfaces of $\mathrm{CuMnAs}$ and $\mathrm{NiO}$ respectively exhibit a series of antiferroically, as well as roughness-sensitive, ferroically ordered, modifications of the surface magnetic dipole moments, via canting or changes in sublattice magnitude, consistent with the bulk ordering of the magnetic multipoles. Our findings demonstrate a universal bulk-boundary correspondance allowing the general prediction of minimal possible surface and interface magnetic modifications, even in non-magnetoelectric materials. Furthermore, it paves the way for more accurate interpretations of a wide variety of surface-sensitive measurements.  
\end{abstract}

\maketitle
\section{Introduction}\label{sec:intro}
\begin{figure}[htp]
    \centering
    \includegraphics{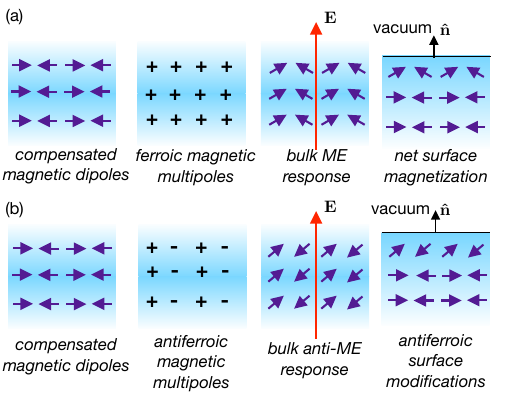}
    \caption{Link between ordering of bulk atomic-site magnetic multipoles, bulk magnetoelectric (ME) effects, and surface magnetic order. (a) Even for surfaces which are parallel to planes with compensated magnetic dipoles in the  bulk (left), ferroic higher-order magnetic multipoles (HOMMs) (second panel) imply a net bulk magnetoelectric response (third panel) and a net induced surface magnetization (right). (b) HOMMs which are antiferroically ordered parallel to the surface plane of interest imply a bulk anti-ME effect, and surface dipolar modifications which maintain the compensation of magnetic dipoles. Fadeout blue color represents continuing periodic bulk, black line in right panels represents vacuum-terminated surface.}
    \label{fig:multipole_ME_surfM} 
\end{figure}

The assumption that magnetic dipole moments on a vacuum-terminated surface or interface maintain the same orientation and magnitude as they would in the bulk is generically inaccurate. Because the creation of a surface leads to an inherent symmetry lowering, additional dipole components are often allowed at a surface despite being forbidden in the bulk. Further, an elegant bulk-boundary correspondence exists between the symmetry of surface dipole magnetic order and that of the magnetoelectric (ME) effect in a bulk material, whereby an applied electric field $\mathbf{E}$ induces a net magnetization $\mathbf{M}$\cite{Astrov1960,Dzyaloshinskii1959}. More concretely, a magnetic crystal with a finite bulk ME response for a particular direction $\mathbf{r}$ of applied electric field will develop the same components of magnetization on a surface perpendicular to $\mathbf{r}$, in the absence of an applied field\cite{Krichevtsov1993,Belashchenko2010,Weber2024}.\\ 
\indent The above argument implies that any bulk antiferromagnet (AFM) which has a nonzero ME response (at linear or higher order) has a corresponding roughness-robust net magnetic dipole moment per unit area (``surface magnetization"), despite the vanishing bulk magnetization\cite{Weber2024} (Fig.~\ref{fig:multipole_ME_surfM}(a)). Such surface magnetization has utility as a directly detectable probe of the underlying AFM bulk N\'{e}el vector\cite{Belashchenko2010,Pylypovskyi2024}, even on surfaces parallel to atomic planes which are magnetically compensated in the bulk (Fig.~\ref{fig:multipole_ME_surfM}(a))\cite{Belashchenko2010,Lapa2020,Weber2024,Wang2014,Lapa2020,Du2023,Pylypovskyi2024}. We showed in Ref.~\citenum{Weber2024} that such ``induced" surface magnetization in ME AFMs occurs when higher-order \textit{local} bulk magnetic multipoles centered on the magnetic sites order ferroically in the unit cell (Fig. \ref{fig:multipole_ME_surfM}(a), second and fourth panels). These atomic-site higher-order magnetic multipoles (HOMMs), which describe magnetic density asymmetries beyond the spherically symmetric magnetic dipole around individual atoms, are established as symmetry indicators for the local bulk ME responses of individual ions, based on the symmetries of their Wyckoff sites\cite{Urru2022,Bhowal2021}, with ferroically ordered magnetic multipoles implying macroscopic ME responses (Fig.~\ref{fig:multipole_ME_surfM}(a)).\\ 
\indent On the other hand, surfaces of AFMs parallel to planes which have antiferroically ordered HOMMs in addition to antiferroically ordered magnetic dipoles (Fig.  \ref{fig:multipole_ME_surfM}(b), second panel) have no net surface magnetization (nor a net ME response for a field along the surface normal). But magnetic dipole order at the surface will \textit{still} change due to the nonzero local HOMMs and the effective electric field of the surface. While the technological utility of antiferroic surface magnetic modifications is less obvious than cases where a net magnetization arises, predicting changes in magnetic order at a surface is crucial for properly interpreting surface-sensitive experiments\cite{Das2015,Trainer2021}. \\
\indent In this work, we show using a combination of symmetry analysis and density functional theory (DFT) that the \textit{minimal} set of surface magnetic dipole modifications, arising  solely from vacuum termination of the periodic bulk, can be universally understood based on the ordering of the bulk atomic-site HOMMs, or equivalently, the \textit{local} ME responses of individual atoms. We further map this local magnetic multipole/local ME response description to an alternative group-theory procedure based on the bulk Wyckoff magnetic point group of the atoms, and the surface orientation of interest. Finally, we use DFT to test our generic arguments for two specific AFMs, topological semimetal $\mathrm{CuMnAs}$ and rock-salt $\mathrm{NiO}$.
\section{Theoretical Background}\label{sec:theory}Most of the background necessary for our analysis was introduced in Ref.\cite{Weber2024}, where we focused exclusively on surfaces of bulk AFMs with a \textit{net} dipole magnetization. Here, we restate the key points and then extend the analysis to allow for the prediction of antiferroic surface magnetic modifications.\\
\indent For ME materials having by definition broken time-reversal symmetry $\Theta$, the induced bulk magnetization along direction $\hat{i}$ due to an applied electric field $\mathbf{E}$ can be written as an expansion in powers of $\mathbf{E}$\cite{Fiebig2005}:
\begin{equation}
    M_i=\alpha_{ij}E_i+\beta_{ijk}E_jE_k+\gamma_{ijkl}E_jE_kE_l+...,
    \label{eq:ME_effect}
\end{equation}
with $\alpha$, $\beta$ and $\gamma$ being the linear, quadratic, and cubic ME response tensors respectively, and summation over repeated indices implied. The various orders of the ME effect can be predicted via symmetry considerations based on ground state properties, namely, the previously mentioned bulk magnetic multipoles, which are the coefficients in a multipole expansion of the interaction energy of a bulk magnetization density $\boldsymbol{\mu}(\mathbf{r})$ with a spatially inhomogenous magnetic field $\mathbf{H}(\mathbf{r})$\cite{Ederer2007}. The coefficient of the first-order term, $\int \boldsymbol{\mu}(\mathbf{r})d^3\mathbf{r}$, gives the magnetic dipole per unit volume, and the next three coefficients, $\mathcal{M}_{ij}=\int r_j\mu_i(\mathbf{r})d^3\mathbf{r}$, $\mathcal{O}_{ijk}=\int r_jr_k\mu_i(\mathbf{r})d^3\mathbf{r}$ and $\mathcal{H}_{ijkl}=\int r_jr_kr_l\mu_i(\mathbf{r})d^3\mathbf{r}$, are HOMMs per unit volume known as the magnetoelectric (ME) multipole, the magnetic octupole, and the magnetic hexadecapole. $\mathcal{M}_{ij}$, $\mathcal{O}_{ijk}$ and $\mathcal{H}_{ijkl}$ have the same symmetry-allowed forms as the linear, quadratic and cubic ME response tensors $\alpha_{ij}$, $\beta_{ijk}$ and $\gamma_{ijkl}$ respectively\cite{Ederer2007,Spaldin2008,Spaldin2013,Urru2022,Weber2024}.\\
\begin{figure}[htp]
    \centering
    \includegraphics{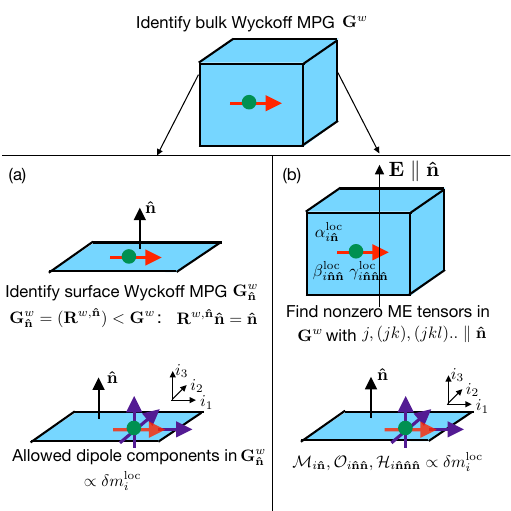}
    \caption{Summary of two possible methods to determine changes in magnetic dipole components (shown in the bottom panels by purple arrows superimposed on the bulk direction and magnitude of the local bulk dipole, indicated with a red arrow) of an individual atom (green circle) at a surface with orientation $(hkl)\perp\mathbf{\hat{n}}$. (a) Method in which symmetry-allowed components of the local magnetic dipole are determined by finding the surface Wyckoff MPG. (b) Method in which the symmetry-allowed bulk ME tensors in $\mathbf{G}^w$ for an electric field along the surface normal, and consequently the nonzero HOMMs, are identified, thus determining the components of the local magnetic dipole at the surface. Note that magnetic octupole $\mathcal{O}_{i\mathbf{\hat{n}}\mathbf{\hat{n}}}$ with $i$ parallel to the bulk direction of magnetic dipole in $\mathbf{G}^w$, corresponding to a change in magnitude along $i$, is always present.}
    \label{fig:surface_symm} 
 \end{figure}
\indent We now make the step from bulk ME effects and bulk magnetic multipoles to the equilibrium magnetic order at a surface. The magnetic point group (MPG) of a two-dimensional unreconstructed surface, whose orientation $(hkl)$ is defined by the unit vector $\mathbf{\hat{n}}$ perpendicular to the surface, is identical to the MPG selected by an electric field applied in the bulk, parallel to $\mathbf{\hat{n}}$~\cite{Weber2024}. This lower-symmetry MPG is defined as the subgroup of bulk MPG operations which leave the direction of the polar vector (either $\mathbf{\hat{n}}$ or $\mathbf{E}$) invariant\cite{Belashchenko2010,Weber2024}. Thus, if the MPG selected by $\mathbf{E}$ in the bulk allows for an induced bulk magnetization, then a surface cleaved such that $\mathbf{\hat{n}}\parallel\mathbf{E}$ will have the same symmetry-allowed components of magnetization in the absence of $\mathbf{E}$.\\
\indent The allowed components of net magnetization in the surface MPG identified via the above formalism reflect the symmetry of ME responses at \textit{all} powers of applied electric field. Due to the mapping $\alpha_{ij}\Leftrightarrow\mathcal{M}_{ij}$, $\beta_{ijk}\Leftrightarrow\mathcal{O}_{ijk}$, $\gamma_{ijkl}\Leftrightarrow\mathcal{H}_{ijkl}$ between each power of ME effect and the corresponding bulk magnetic multipole, determining which HOMMs are ferroically ordered in the unit cell immediately tells us which orders of the bulk ME effect are reflected in the symmetry-allowed components of surface magnetization.\\ 
\indent We now extend the above analysis by noting that even if a bulk MPG forbids any net ME response (that is, there is no HOMM which is ferroically ordered throughout the unit cell), or equivalently the surface MPG has no roughness-robust surface magnetization, individual atomic-site HOMMs can still be nonzero, as long as the corresponding ME tensor is allowed by the lower-symmetry MPG of the Wyckoff site of the atom\cite{Thole2018,Thole2018a}. The bulk Wyckoff MPG for a given atomic site is defined as the subgroup of bulk operations that map the site onto itself. Recent theoretical work\cite{Verbeek2023} demonstrated that an antiferroic ordering of HOMMs leads to anti-magnetoelectric effects, whereby an applied electric field induces changes in individual magnetic moments which cancel across the unit cell (Fig. \ref{fig:multipole_ME_surfM}(b), third panel).\\
\indent Starting from the bulk Wyckoff MPG $\mathbf{G}^w$ for the magnetic atom in question, we can identify the \textit{surface} Wyckoff MPG $\mathbf{G}_{\mathbf{\hat{n}}}^w$ for a specific surface orientation characterized by $\mathbf{\hat{n}}$ as the subset of operations in $\mathbf{G}^w$ which leave the direction of $\mathbf{\hat{n}}$ invariant:
\begin{equation}
    \mathbf{R}_i\in \mathbf{G}_{\mathbf{\hat{n}}}^w\iff \mathbf{R}_i\mathbf{\hat{n}}=\mathbf{\hat{n}},
\label{eq:surf_wyckoffMPG}
\end{equation}
where $\mathbf{R}_i$ is the $i^{\mathrm{th}}$ element of the bulk Wyckoff MPG $\mathbf{G}^w$. Then, we examine which components $(m_x,m_y,m_z)$ are allowed in $\mathbf{G}_{\mathbf{\hat{n}}}^w$ to deduce the possible modifications of the surface magnetic dipole moment compared to its symmetry-allowed components in $\mathbf{G}^w$ (Fig. \ref{fig:surface_symm}(a)).\\
\indent Alternatively, we can determine all symmetry-allowed ME tensors within $\mathbf{G}^w$, and from this deduce the atomic-site magnetic multipoles which have nonzero values for the components with all spatial indices $r_j,r_k,r_l...\parallel\mathbf{\hat{n}}$ (since these are the components mapping to the local bulk ME response for $\mathbf{E}\parallel\mathbf{\hat{n}}$). The full set of these HOMMs tells us the allowed components of induced dipole magnetization for the individual atoms at a surface (Fig. \ref{fig:surface_symm}(b)).\\
\indent Before moving to specific examples, we note that the allowed magnetic dipole components for an atomic site at a surface based on the analysis above in general depend on the specific symmetries maintained in $\mathbf{G}^w_{\mathbf{\hat{n}}}$. There is however one set of nonzero HOMMs, and a corresponding local magnetic modification, that is \textit{always} present at a surface. Recall that the bulk Wyckoff MPG must allow for at least one equilibrium component of local magnetic dipole $m_{i}$ in a magnetically ordered material. Interestingly, one can prove (see Supplementary Material) that if a local magnetic dipole is symmetry-allowed along direction $i$ for $\mathbf{G}^w$, then all second-order magnetic octupoles with diagonal spatial indices, that is $\int d^3\mathbf{r}\mu_i(\mathbf{r})r_jr_{k=j}$ for $j=(x,y,z)$ are also symmetry-allowed.\\
\indent This means that for both ferromagnets and AFMs, any magnetically ordered site must also have a local second-order ME response, with the induced magnetization parallel to the bulk direction of the dipole, for \textit{any} direction of applied electric field. It also follows that for any surface orientation, the magnetic dipole of a surface atom has a symmetry-allowed magnetization change parallel to its direction in the bulk, which manifests as a change in its dipole moment magnitude. This omnipresent coexistence of atomic-site magnetic dipoles $m_i$ and magnetic octupoles $\mathcal{O}_{ijj}$ is consistent with the generic magnitude changes of dipole moments at surfaces and interfaces, often by tenths of a $\mu_B$, with respect to their bulk values\cite{Eriksson1991,Siegmann1992}.\\
\indent We summarize the above-described analysis for identifying all allowed magnetic dipole components, for individual surface atoms in Fig.~\ref{fig:surface_symm}. We now apply these generic arguments to the analysis of surfaces of two specific AFMs: $\mathrm{CuMnAs}$ and $\mathrm{NiO}$.
\section{CuMnAs}\label{sec:cumnas}
\begin{figure}
    \centering
    \includegraphics{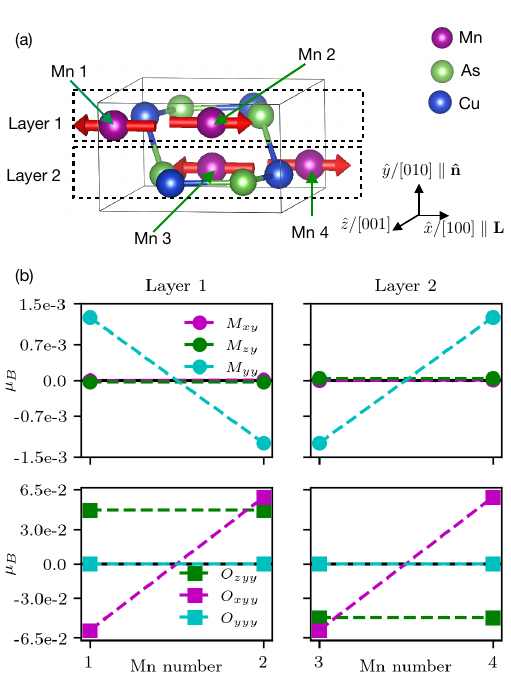}
    \caption{Crystal structure and multipoles for $\mathrm{CuMnAs}$. (a) Bulk crystal structure of orthorhombic $\mathrm{CuMnAs}$. Each atomic layer parallel to the $(010)$ surface of interest contains two oppositely-oriented $\mathrm{Mn}$ sublattices (the bulk magnetic unit cell contains two layers, a total of four $\mathrm{Mn}$. (b) Calculated bulk magnetic multipoles for the four $\mathrm{Mn}$ atoms, with the bulk dipole orientations shown in (a). Left column is layer 1, right column is layer 2. Labeling of $\mathrm{Mn}$ atoms is the same as in (a). Top: ME multipoles. Bottom: magnetic octupoles. The dashed lines serve as visual aids to connect the set of HOMMs, and are not meant to imply a linear relation between the atoms in each layer.}
    \label{fig:CuMnAs_symm} 
\end{figure}
Orthorhombic $\mathrm{CuMnAs}$ is a semimetal with a collinear AFM order. It crystallizes in space-group \textit{Pnma}. In the last decade it has garnered interest based on theoretical proposals that it hosts gapless topological Dirac fermions if the N\'{e}el vector is oriented along the $[001]$ direction, whereas the Dirac crossings gap out for other N\'{e}el vector orientations\cite{Tang2016,Smejkal2017}. Furthermore, like its better-known tetragonal counterpart, the efficient reorientation of the N\'{e}el vector via current-induced spin-orbit torque should be possible based on symmetry arguments\cite{Smejkal2017}, thus enabling an electrically controlled tuning of the topology.\\
\indent We focus on the $(010)$ surface of orthorhombic $\mathrm{CuMnAs}$, which is parallel to bulk atomic layers having one $\mathrm{Mn}$ atom of each sublattice; thus, with the bulk collinear AFM order, these layers have completely compensated magnetic dipoles (Fig. \ref{fig:CuMnAs_symm}(a)). We mention here that there is some discrepancy between experiment and theory regarding the ground-state orientation of the bulk N\'{e}el vector\cite{Zhang2017,Emmanouilidou2017}. Our DFT calculations indicate that the $[100]$ orientation of the N\'{e}el vector is lower in energy by $0.05$ ($0.1$) $\mathrm{meV}$ per formula unit than the $[010]$ ($[001]$) orientation, also in agreement with prior first-principles studies\cite{Smejkal2017}. Thus for concreteness we perform the following analysis assuming that $\mathbf{L}\parallel[100]$, as in Fig. \ref{fig:CuMnAs_symm}(a).\\
\indent First we use the method in Fig. \ref{fig:surface_symm}(a) to identify the symmetry-allowed magnetic dipole components for individual $\mathrm{Mn}$ atoms on the $(010)$ surface by finding their surface Wyckoff MPG. With $\mathbf{L}\parallel[100]$, the full bulk magnetic space group (MSG) of $\mathrm{CuMnAs}$ is $Pnm'a$ [62.444]\footnote{Actually, due to its broken inversion and time-reversal symmetries, this bulk MSG allows for a \textit{net} off-diagonal linear ME response in $\mathrm{CuMnAs}$ for an electric field along $\hat{x}/[100]$ (induced magnetization along $\hat{z}/[001]$) or $\hat{z}/[001]$ (induced magnetization along $\hat{z}/[100]$). However, for an electric field along $\hat{y}/[010]$, there is no net ME effect at linear or higher order, and equivalently, no net, roughness-robust $(010)$ surface magnetization}. The $\mathrm{Mn}$ moments sit at the Wyckoff site $4c$, and with this MSG the Wyckoff site corresponds to the bulk Wyckoff point group $\mathbf{G}^w=m'$. $m'$ contains two symmetry operations; the trivial identity $\mathcal{E}$ and a mirror plane $\sigma_{[010]}$ perpendicular to $\hat{y}/[010]$ coupled to time reversal, $\sigma'_{[010]}$. These operations allow for bulk magnetic dipole components parallel to $\hat{x}/[100]$ and $\hat{z}/[001]$ on the individual Wyckoff sites. Although a local dipole component along $\hat{z}$ is already symmetry-allowed in the bulk, according our DFT calculations, the magnetoanisotropy energy of $\mathrm{CuMnAs}$ constrains the bulk N\'{e}el vector to point purely along $\hat{x}/[100]$.\\
\indent Now to find the $(010)$ surface Wyckoff MPG, we consider which operations in $m'$ leave the surface normal $\mathbf{\hat{n}}$ invariant. Under the operation $\sigma'_{[010]}$, spatial coordinates transform as $(x,y,z)\rightarrow(x,-y,z)$. Thus since $\sigma'_{[010]}$ flips the sign of $\mathbf{\hat{n}}\parallel[010]$, it is forbidden in the surface Wyckoff MPG and we are left with the trivial group having only the identity, $\mathbf{G}^w_{\mathbf{\hat{n}}}=\mathbf{1}$, which allows for a net magnetic dipole along all three Cartesian directions, in contrast to the bulk Wyckoff MPG which forbids any local dipole component along $\hat{y}/[010]$. This indicates that on the $(010)$ surface, in addition to an atomic-site magnetic dipole $m_x$ along the bulk N\'{e}el vector direction, we can also expect cantings along $\hat{y}/[010]$ and $\hat{z}/[001]$ of the individual $\mathrm{Mn}$ surface dipole moments.\\
\indent We now use the alternative method of analysis (Fig. \ref{fig:surface_symm}(b)) where, instead of finding the surface Wyckoff MPG, we identify the symmetry-allowed ME tensors within the bulk Wyckoff MPG $m'$ which have nonzero components for an electric field along $\hat{y}/[010]$. Adopting a Cartesian coordinate system where the $\hat{y}$ coordinate (perpendicular to the mirror plane) is parallel to $\mathbf{\hat{n}}$, and $\hat{x}$ is parallel to $\mathbf{L}$, the local linear ME response tensor for $\mathbf{G}^w$ takes the form
\begin{equation}
    \alpha^{\mathrm{loc}}=\begin{pmatrix}
    \alpha_{xx} & 0 & \alpha_{xz}\\
    0 & \alpha_{yy} & 0\\
    \alpha_{zx} & 0 & \alpha_{zz}
    \end{pmatrix}.
    \label{eq:ME_multipoles_CuMnAs}
\end{equation}
\indent From this, we can immediately conclude that  atomic-site ME multipoles $\mathcal{M}^{\mathrm{loc}}_{yy}$ on individual atoms are nonzero, and thus for the relevant surface normal direction $\mathbf{\hat{n}}\parallel\hat{y}$ this means that $\mathrm{Mn}$ moments at the $(010)$ surface have a symmetry-allowed out-of-plane canting. We can compare Eq. \ref{eq:ME_multipoles_CuMnAs} to the form of the linear ME tensor for the \textit{full} bulk MPG of orthorhombic $\mathrm{CuMnAs}$, which is $mm'm$. $\alpha$ in this MPG has vanishing components on the diagonal. Thus, the atomic-site ME multipoles $\mathcal{M}^{\mathrm{loc}}_{yy}$ must be ordered antiferroically across the four $\mathrm{Mn}$ atoms in the unit cell. This is exactly what we see in the top panel of Fig. \ref{fig:CuMnAs_symm}(b), where we have explicitly calculated using DFT the three atomic-site ME multipoles with spatial indices along $\hat{y}$ for the orientation and domain of bulk magnetic dipoles shown in Fig. \ref{fig:CuMnAs_symm}(a). As expected, $\mathcal{M}^{\mathrm{loc}}_{xy}$ and $\mathcal{M}^{\mathrm{loc}}_{zy}$ vanish, while $\mathcal{M}^{\mathrm{loc}}_{yy}$ values on $\mathrm{Mn}1$ and $\mathrm{Mn}4$ are equal and opposite to those on $\mathrm{Mn}2$ and $\mathrm{Mn}3$. From this we know that the $\mathrm{Mn}$ atoms for layers near a vacuum-terminated $(010)$ surface (for example, $\mathrm{Mn}1$ and $\mathrm{Mn}2$) will experience antiferroically ordered out-of-plane cantings.\\
\indent Now we examine the components of the second-order ME tensor in $\mathbf{G}^w$ which correspond to an electric field along $\hat{y}$.  In MPG $m'$ the symmetry-allowed components $\beta_{iyy}$ are $\beta_{xyy}$ and $\beta_{zyy}$, with $\beta_{yyy}$ constrained to zero. This contrasts with the case of the full bulk MPG $mm'm$, for which the 2nd order ME tensor is null. Thus, the corresponding individual magnetic octupoles $\mathcal{O}^{\mathrm{loc}}_{xyy}$ and $\mathcal{O}^{\mathrm{loc}}_{zyy}$ are nonzero but, like the $\mathcal{M}^{\mathrm{loc}}_{yy}$ ME multipoles, are antiferroically ordered in the unit cell. We confirm this by explicitly calculating the atomic-site magnetic octupoles in the bottom panel of Fig.\ref{fig:CuMnAs_symm}(b). We see that the $\mathcal{O}^{\mathrm{loc}}_{yyy}$ octupoles vanish for all $\mathrm{Mn}$ atoms, whereas both $\mathcal{O}^{\mathrm{loc}}_{zyy}$, corresponding to a local in-plane canting of $(010)$ $\mathrm{Mn}$ surface moments perpendicular to the N\'{e}el vector, and $\mathcal{O}_{xyy}$, corresponding to a change in magnitude at the $(010)$ surface parallel to $\mathbf{L}$, are nonzero and antiferroically ordered on the four $\mathrm{Mn}$ atoms in the unit cell. Thus, the alternative method for local surface magnetic modifications based on analysis of the ME tensors of the bulk Wyckoff MPG brings us to the same conclusion as directly finding the surface Wyckoff MPG. That is, at the $(010)$ surface individual $\mathrm{Mn}$ dipole moments have symmetry-allowed components along all three Cartesian directions.\\
\indent There is one difference of note between the pattern of antiferroic ordering of the $\mathcal{O}^{\mathrm{loc}}_{xyy}$ and $\mathcal{O}^{\mathrm{loc}}_{zyy}$ octupoles, which is not immediately apparent from our first method of analysis. Fig. \ref{fig:CuMnAs_symm}(b) shows that the $\mathcal{O}^{\mathrm{loc}}_{xyy}$ octupoles, like the $\mathcal{M}^{\mathrm{loc}}_{yy}$ ME multipoles, are antiferroically ordered in the two $\mathrm{Mn}$ atoms in each layer parallel to $(010)$, such that the changes in magnitude of the magnetic dipoles for the two sublattices in each layer will be of opposite sign. On the other hand, the $\mathcal{O}_{zyy}$ octupoles are positive for both atoms $\mathrm{Mn}1$ and $\mathrm{Mn}2$ in layer one, and equal magnitude but negative for both atoms $\mathrm{Mn}3$ and $\mathrm{Mn}4$ in layer 2. This means that $(010)$ $\mathrm{CuMnAs}$ in fact has a nonzero, \textit{roughness-sensitive} induced surface magnetization along $\hat{z}/[001]$, which will be finite in the atomically flat limit, but in the presence of atomic steps at the surface, will average macroscopically to zero\footnote{Note that we would have arrived at the same conclusion using the group-theory formalism described in Ref. [5] where we identify roughness-sensitive surface magnetization by identifying the full bulk surface magnetic \textit{space group}, only including operations involving translations parallel to the surface, and then checking for additional operations in the bulk magnetic space group which flip the direction of surface magnetization and involve a translation perpendicular to the surface.}.\\
\begin{figure}[htp]
    \centering
    \includegraphics{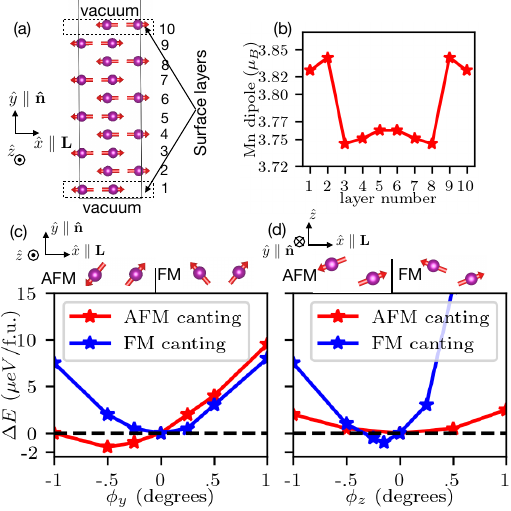}
    \caption{(a) $(010)$-oriented slab of $\mathrm{CuMnAs}$ consisting of ten layers of two $\mathrm{Mn}$ atoms each. $\mathrm{Cu}$ and $\mathrm{As}$ atoms removed for ease of viewing. (b) Magnitude of $\mathrm{Mn}$ sublattice dipole moments along the bulk N\'{e}el vector direction $\hat{x}/[100]$ as a function of layer number. (c) Change in energy with respect to zero canting for rotating surface moments out-of-plane along $\hat{y}/[010]$. The cartoons above the plot show the symmetry-allowed AFM canting (left) and symmetry-disallowed FM canting (right) for one surface layer when looking along the surface. (d) Change in energy with respect to zero canting for rotating surface moments in-plane along $\hat{z}/[001]$. The cartoons above the plot show the symmetry-disallowed AFM canting (left) and symmetry-allowed FM canting (right) for one surface layer when looking down on the surface.} 
    \label{fig:CuMnAs_E} 
\end{figure}
\indent While symmetry considerations tell us which types of local magnetic modifications \textit{can} occur, they do not prove that they are energetically favorable with respect to the bulk collinear AFM order. To demonstrate that our symmetry predictions actually manifest for the $(010)$ surface of orthorhombic $\mathrm{CuMnAs}$, we perform constrained magnetic total energy calculations within DFT+U, including spin-orbit coupling (see appendix for computational details). Fig.~\ref{fig:CuMnAs_E}(a) shows the vacuum-terminated slab configuration used in the calculations, consisting of ten layers or five unit cells. Fig.~\ref{fig:CuMnAs_E}(b) shows the magnetic dipoles magnitude (the two oppositely-oriented dipole moments in each layer have equal magnitudes as required by symmetry) as a function of layer number. We see that, consistently with the nonzero $\mathcal{O}^{\mathrm{loc}}_{xyy}$ octupoles, the moments on the two top and bottom layers closest to vacuum change on the order of $0.1\mu_B$ with respect to the moments in the central, bulk-like layers.\\
\indent Next, in Figs.~\ref{fig:CuMnAs_E}(c) and (d) we examine the energetics of the noncollinear out-of-plane (along $[010]$) and in-plane (along $[001]$) canting respectively of $(010)$ surface $\mathrm{Mn}$ moments. We constrain the directions of $\mathrm{Mn}$ dipole moments in the central bulk-like layers 2-9 to the collinear $[100]$ direction, and systematically cant the moments on the top and bottom surface layers (see appendix for details). For both directions of rotation, we check the change in energy for both antiferroic and ferroic canting of the two moments in a single surface layer. In Fig.~\ref{fig:CuMnAs_E}(c), when canting the surface moments out of plane, we see an energetic stabilization for a finite value of $\phi_y$ (about $-0.5$ degrees for the selected bulk AFM domain; note that switching the bulk domain results in the opposite angle of energetically favored out-of-plane rotation) when we cant the $\mathrm{Mn}$ moments in a single surface layer in opposite directions (red curve). This is in agreement with the in-plane antiferroic pattern of $\mathrm{M}^{\mathrm{loc}}_{yy}$ ME multipoles (Fig. \ref{fig:CuMnAs_symm}(b)). In contrast, a symmetry-disallowed, ferroically ordered pattern of out-of-plane canting (blue curve) leads to a symmetric increase in energy with respect to $\phi_y=0$ for either positive or negative out-of-plane canting.\\
\indent We see the exact opposite for in-plane canting along $[001]$ in Fig.~\ref{fig:CuMnAs_E}(d). In this case, an antiferroic pattern of canting leads to a symmetric increase in energy for either sign of rotation of the N\'{e}el vector. On the other hand, a ferroic pattern of canting leads to an asymmetric profile about $\phi_z=0$ with an energy stabilization at a small angle of $0.15$ degrees. This implies a nonzero, roughness-sensitive surface magnetization along $[001]$ of about $-0.01\mu_B$ per $\mathrm{Mn}$ atom, in agreement with the in-plane ferroic, and out-of-plane antiferroic, ordering of the  $\mathcal{O}^{\mathrm{loc}}_{zyy}$ octupoles.\\ 
\indent We now give one final material example, rock-salt $\mathrm{NiO}$, which in contrast to $\mathrm{CuMnAs}$, is an insulating AFM with \textit{no} symmetry-allowed induced surface magnetization.
\section{NiO}\label{sec:nio}
\begin{figure}
    \centering
    \includegraphics{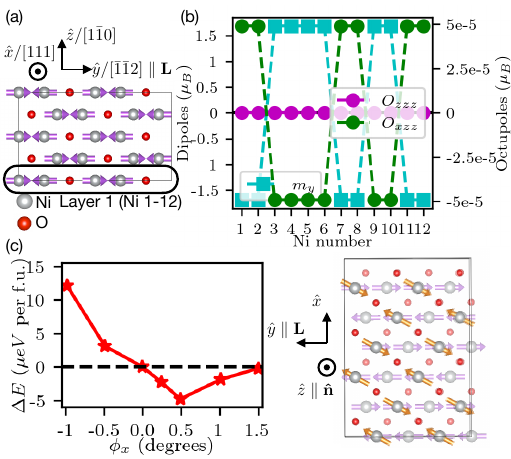}
    \caption{(a) Bulk supercell of $\mathrm{NiO}$ oriented such that it tiles the $(1\bar{1}0)$ surface. Each layer parallel to the surface contains twelve $\mathrm{Ni}$ moments (example bottom layer is circled in black), and the layers are magnetically compensated in bulk. Some atoms along the $[111]$ direction (into the page) deleted for clarity of plot. (b) Bulk magnetic dipoles and magnetic octupoles $\mathcal{O}_{zzz}$ and $\mathcal{O}_{xzz}$ corresponding to a surface normal along $\hat{z}/[1\bar{1}0]$ for the twelve $\mathrm{Ni}$ moments in a single layer parallel to $(1\bar{1}0)$. Labeling $1$-$12$ of the  moments is taken directly from the structure file. (c) Constrained magnetic DFT calculations showing change in energy with respect to zero canting for the symmetry-allowed antiferroic canting along $\hat{x}/[111]$ (symmetry-allowed). Righthand cartoon shows the in-plane rotation $\phi_x$ of the dipole moments in top surface layer (orange arrows) performed in the constrained magnetic calculations, corresponding to $\mathrm{Ni}$ 1 through 12 in (b). Faded purple arrows are $\mathrm{Ni}$ dipole moments of lower layers.}
    \label{fig:NiO} 
\end{figure}
Having worked through the example of $(010)$ $\mathrm{CuMnAs}$ in some detail, we will more briefly describe the analysis of surface magnetic modifications in $\mathrm{NiO}$. $\mathrm{NiO}$ is a rock-salt structure, insulating AFM, with the N\'{e}el vector oriented along the $[\bar{1}\bar{1}2]$ cubic direction. $\mathrm{NiO}$ has attracted interest for decades due to its myriad energy-related applications in batteries, chemical sensors, catalytic processes and more\cite{SaniGarbaDanjumma2019,DiGirolamo2020}. Notably for our focus, $\mathrm{NiO}$'s bulk magnetic space group (MSG), $C_c2/c$ [15.90], is a so-called type-IV MSG, that is, it contains time reversal $\Theta$ coupled to one or more translations as symmetry elements. For $\mathrm{NiO}$ in particular, any two oppositely oriented $\mathrm{Ni}$ magnetic moments in the primitive unit cell are connected by one of the two symmetry operations $(\Theta|0,0,1/2)$ and $(\Theta|1/2,1/2,1/2)$ in $C_c2/c$ (with the translations here written in the primitive monoclinic, rather than cubic, basis). Because every magnetic multipole is odd under $\Theta$, it directly follows that \textit{all} HOMMs follow exactly the bulk antiferromagnetic pattern of the magnetic dipoles. Thus, a surface of $\mathrm{NiO}$ which is parallel to magnetically compensated planes in bulk can never have a net, induced magnetization; all surface magnetic modifications on nominally compensated surfaces \textit{must} be antiferroic.\\
\indent We show this explicitly for the $(1\bar{1}0)$ surface (now again adopting the cubic rock-salt basis for convenience). Note that we already discussed this surface in Ref.~\citenum{Weber2024} in the context of a ``null" case with no net surface magnetization. We now focus instead on possible antiferroic magnetic modifications at the $(1\bar{1}0)$ surface. Fig. \ref{fig:NiO}(a) shows a bulk supercell of $\mathrm{NiO}$ with in-plane orthogonal lattice vectors along $\hat{x}/[111]$ and $\hat{y}/[\bar{1}\bar{1}2]$ directions, and out-of-plane lattice vector along the $[1\bar{1}0]$ surface normal of interest. Again, note that the three indices for each lattice vector of the supercell refer to the $[100]$, $[010]$ and $[001]$ directions respectively of the rock-salt cubic structure. The bulk N\'{e}el vector $\mathbf{L}\parallel[\bar{1}\bar{1}2]$ lies in-plane, and each layer of 12 $\mathrm{Ni}$ atoms parallel to the $(\bar{1}10)$ surface has six positive, and six negatively oriented dipole moments, and is thus magnetically compensated.\\
\indent Having worked through both methods depicted in Fig. \ref{fig:surface_symm} in detail for the example of $\mathrm{CuMnAs}$, for $\mathrm{NiO}$ we focus on the method in Fig. \ref{fig:surface_symm}(b) where we identify the allowed magnetic $(1\bar{1}0)$ surface modifications by analyzing the local ME tensors for the bulk $\mathbf{G}^w$. The $\mathrm{Ni}$ moments sit at bulk Wyckoff position $4c$, and for the bulk MSG $C_c2/c$ this corresponds to Wyckoff MPG $\mathbf{G}^w=2'/m'$. $2'/m'$ contains inversion as a symmetry. Thus, because the linear ME tensor, as well as the ME multipoles, are odd under $\mathcal{I}$, the linear ME tensor must be null for this MPG. So we instead examine the symmetry-allowed components of the second-order ME tensor $\beta$. Rotating $\beta$ such that $\hat{x}$, $\hat{y}$ and $\hat{z}$ are along the $[111]$, $[\bar{1}\bar{1}2]$ and $[1\bar{1}0]$ crystallographic directions respectively, as in Fig.~\ref{fig:NiO}(a), the quadratic ME tensor elements corresponding to an electric field along $\hat{z}$, that is $\beta_{izz}$, are nonzero for $\beta_{xzz}$ and $\beta_{yzz}$ but zero for $\beta_{zzz}$.\\
\indent Fig. \ref{fig:NiO}(b) shows the calculated corresponding bulk magnetic octupoles $\mathcal{O}^{\mathrm{loc}}_{zzz}$ and $\mathcal{O}^{\mathrm{loc}}_{xzz}$ for the 12 $\mathrm{Ni}$ moments in a layer parallel to $(1\bar{1}0)$, together with the magnetic dipoles. As expected, the $\mathcal{O}^{\mathrm{loc}}_{zzz}$ octupoles are zero, while the $\mathcal{O}^{\mathrm{loc}}_{xzz}$ octupoles corresponding to an in-plane canting of the surface $\mathrm{Ni}$ moments are antiferroically ordered and follow exactly the pattern of the sign of the dipoles (though with opposite sign). Thus, we expect antiferroic in-plane cantings to be energetically stable, whereas both antiferroic and ferroic out-of-plane cantings along $\hat{z}/[1\bar{1}0]$ are symmetry-disallowed. Note that for clarity of the plot we do not show the $\mathcal{O}_{yzz}$ octupoles corresponding to a change in magnitude along the direction of the bulk N\'{e}el vector, but have confirmed that as expected these octupoles are nonzero and follow the dipole ordering pattern. We also see a corresponding change in the surface magnetic moment magnitudes in our DFT calculations.\\
\indent In Fig. \ref{fig:NiO}(c) we again use constrained-magnetic DFT calculations to demonstrate that the symmetry-allowed antiferroic canting along the in-plane $[111]$ direction is energetically favorable. We focus exclusively on antiferroic canting of oppositely oriented $\mathrm{Ni}$ sublattices, as we already addressed the absence of ferroic canting and a corresponding induced surface magnetization, in Ref.~\citenum{Weber2024}. As shown in Fig. \ref{fig:NiO}(c) we confirm a clear energetic preference for a nonzero, in-plane canting (corresponding to $\phi_x\approx0.5^{\circ}$) ordered antiferroically in the surface plane. When we enforce symmetry-disallowed out-of-plane canting on the other hand (corresponding to $\phi_z$ rotation), the total energy oscillates noisily with respect to its value at $\phi_z=0$ and does not exhibit a systematic energy lowering for finite canting\footnote{This direction of symmetry-disallowed rotation for $\mathrm{NiO}$ was particularly difficult to converge compared to all other constrained-energy calculations we describe, and the random, small increases and decreases in total energy we observed for different rotation angles are within the noise of the. calculation Thus, we chose to leave the results out of the figure to avoid over-interpretation by the reader}.\\
\indent Before moving to broader conclusions, we briefly summarize our material-specific findings for both $\mathrm{CuMnAs}$ and $\mathrm{NiO}$. We have shown that, despite having no \textit{net} ME response for an electric field along $[010]$, and no corresponding roughness-robust magnetization on its $(010)$ surface, semimetallic, orthorhombic $\mathrm{CuMnAs}$ should exhibit \textit{local} changes in $\mathrm{Mn}$ magnetic dipole moments at its $(010)$ surface along all three Cartesian directions. By examining the relative signs of bulk HOMMs on $\mathrm{Mn}$ atoms, we predict antiferroic out-of-plane cantings of the $\mathrm{Mn}$ $(010)$ surface dipole moments, antiferroic changes in surface dipole magnitudes, and a roughness-sensitive $[001]$-oriented surface magnetization. We confirm these symmetry-based predictions by constrained magnetic DFT calculations. The induced changes we find in the local dipole moments, between $0.1$ to $0.01$ $\mu_B$ per $\mathrm{Mn}$ depending on the cartesian component, are over an order of magnitude larger than typical bulk induced moments for insulating ME AFMs subjected to an applied electric field\cite{Verbeek2023,Iniguez2008}. Thus, the physical consequences of such surface-symmetry-induced magnetic modifications, however objectively small, may be non-negligible. Further explorations of the consequences of these surface magnetic dipole changes for the topological surface states of $\mathrm{CuMnAs}$ in particular could be interesting.\\
\indent We have also analyzed the surface-symmetry-induced magnetic dipole modifications of the $(1\bar{1}0)$ surface of $\mathrm{NiO}$. In contrast to our primary material example of $\mathrm{CuMnAs}$ whose ordering of HOMMs corresponds to both ferroic and antiferroic changes in its surface magnetic dipoles, the type-IV magnetic space group of $\mathrm{NiO}$ constrains all HOMMs to follow the pattern of its bulk magnetic dipoles. In general therefore, the allowed magnetic modifications for a given surface of $\mathrm{NiO}$ will always be ferroic (antiferroic) if the surface is parallel to uncompensated (compensated) layers of bulk dipoles. Further investigations of the possible ramifications of these magnetic modifications for $\mathrm{NiO}$, both for experimental characterizations and for its myriad possible applications, are desirable.
\section{Conclusions}\label{sec:conclusions} In this work we have demonstrated that the correspondence between bulk ME responses and surface magnetic order extends beyond the cases where we have a net surface magnetization. Rather, by analyzing the symmetry-allowed \textit{local} ME responses on individual magnetic ions, we can universally determine how magnetic dipole moments at a surface are modified with respect to their bulk ordering even in cases where no net surface magnetization, or net bulk ME responses, exist. This formalism serves as an important starting point for predicting qualitatively the realistic magnetic reordering at surfaces, since it identifies the \textit{minimal} set of symmetry-allowed magnetic changes that occur even for ideal surfaces. We emphasize that even when further symmetry lowering occurs with respect to the pristine bulk-terminated surface, via for example reconstructions, the nonzero local dipole components allowed by the MPG of the higher-symmetry, unreconstructed surface will persist. On top of these minimal surface magnetic changes that are predicted directly from the bulk-boundary correspondence described in this  manuscript, one can then predict \textit{additional} changes from the bulk magnetic order in a similar manner by identifying the subgroup of the pristine surface Wyckoff MPG that describes the reconstructed surface. Going forward, it will be important to better understand the material properties (atomic elements, bulk symmetries, orbital composition) that influence the quantitative magnitudes of these symmetry-allowed modifications, in order to determine the cases in which these surface magnetic reorderings will be most experimentally relevant. We hope our work stimulates further discussions in the theoretical community, and can eventually lead to more nuanced interpretations of surface-sensitive experimental probes, such as spin-polarized STM, ARPES and beyond.
\section{Acknowledgements}
The authors thank Claude Ederer for useful discussions. This
work was funded by the ERC under the European Union’s
Horizon 2020 research and innovation program with Grant
No. 810451 and by ETH Z\"{u}rich. AU acknowledges support
from the Abrahams Postdoctoral Fellowship of the Center for Materials Theory at Rutgers University. Computational resources
were provided by ETH Z\"{u}rich's Euler cluster.
\section{Appendices}
\subsection{Computational Details}
For our density-functional calculations, we employ the Vienna \textit{ab-initio} simulation package (VASP)\cite{Kresse1996} with a generalized gradient approximation (GGA) using the Perdew-Burke-Ernzerhof (PBE) functional\cite{Perdew1996} and projector augmented wave method (PAW)\cite{Blochl1994}. For our study of $\mathrm{CuMnAs}$, we use the standard VASP PAW pseudopotentials with the following valence electron configurations: $\mathrm{Mn}$ $(3d^{6}4s^1)$, $\mathrm{Cu}$ $(3d^{10}4s^1)$, and $\mathrm{As}$ $(4s^{2}4p^{3})$. For calculations of $\mathrm{NiO}$, we use the following valence electron configurations for the PAW pseudopotentials: $\mathrm{Ni}$ $(3p^{6}3d^{9}4s^{1})$, and $\mathrm{O}$ $(2s^22p^4)$. In order to approximately account for the localized nature of the unpaired $\mathrm{Ni}$ $d$ electrons in $\mathrm{NiO}$, we add a Hubbard $U$ correction using the rotationally invariant method by Dudarev \textit{et al.}\cite{Anisimov1997,Dudarev1998}. We set $U=4.6$ $\mathrm{eV}$ for consistency with our previous DFT+U work on $\mathrm{NiO}$~\cite{Weber2024}. Due to the semimetallic, itinerant nature of orthorhombic $\mathrm{CuMnAs}$, we do not employ a Hubbard $U$ correction for this compound. In both cases we use a Gaussian smearing with width of $0.05$ ($0.02$) $\mathrm{eV}$ for $\mathrm{CuMnAs}$ ($\mathrm{NiO}$) respectively.\\
\indent For both structures, we first relax the lattice constants and atomic positions for the bulk magnetic unit cells. To study magnetic order at the $(010)$ ($(1\bar{1}0)$) surfaces of $\mathrm{CuMnAs}$ ($\mathrm{NiO}$) we construct 10-layer (4-layer) slabs with $20\mathrm{\AA}$ of vacuum perpendicular to the surfaces. In both cases, we ensure that the layer-projected density of states for layers in the centers of the slab matches closely the bulk density of states, thus ensuring we have both ``bulk-like" and ``surface-like" regions, while still keeping the slabs to a computationally tractable size. We fix the lattice parameters to the relaxed bulk  values, corresponding to orthogonal in-plane lattice constants of $7.31\mathrm{\AA}$ and $6.58\mathrm{\AA}$ for the $(010)$ surface of $\mathrm{CuMnAs}$ and in-plane lattice constants of $14.62\mathrm{\AA}$ and $10.34\mathrm{\AA}$ for the $(1\bar{1}0)$ surface of $\mathrm{NiO}$. We then allow the atomic positions in the slabs to relax until forces on all atoms are less than $0.01$ $\mathrm{eV}/\mathrm{\AA}$. For both compounds we use a kinetic energy cutoff of $800$ $\mathrm{eV}$ for our plane-wave basis set, and we select a Gamma-centered k-point mesh of $11\times11\times1$ for the $(010$ $\mathrm{CuMnAs}$ slab and $4\times6\times1$ for the $(1\bar{1}0)$ $\mathrm{NiO}$ slab. For all relaxations of both bulk and slab structures, we employ spin-polarized, collinear DFT using the ground-state bulk AFM order and neglecting spin-orbit coupling (SOC).\\
\indent For the calculations of total energy as function of rotation angle for surface magnetic dipole moments, we include SOC self-consistently and furthermore use constrained magnetic calculations to ensure that all magnetic dipole moments remain exactly along the directions we initialize them in. The theory and implementation of constrained DFT for noncollinear magnetism are described in detail in Ref.~\cite{Ma2015}. Briefly, constrained magnetism works by introducing a penalty energy $E_p$ whose weight is governed by the Lagrange multiplier $\lambda$ at each magnetic site. $E_p$ drives up the total energy for any deviation of magnetic dipole moments from their specified directions. For our calculations, we use $\lambda=20$ and $\lambda=10$ for $\mathrm{CuMnAs}$ and $\mathrm{NiO}$ respectively. These values are sufficient to fix all magnetic dipole moments along the desired directions while keeping $E_p$ at least an order of magnitude below the total energy differences between different rotation angles.\\
\indent As a subtle detail, we note that for the cases where we cant surface magnetic dipole moments according to the symmetry-allowed pattern dictated by the HOMMs, we cant dipole moments on both top and bottom surfaces of our slab (we have also confirmed that an energy lowering occurs for canting a single surface). For a given sign and magnitude of rotation angle for the top surface, we determine symmetry-equivalent directions of rotation for the bottom surface based on how the relevant magnetic multipoles behave under inversion symmetry. For concreteness, say that for a surface lying in the $(x,y)$ plane with the  bulk N\'{e}el vector in-plane we have a symmetry-allowed AFM canting along the out-of-plane $\hat{z}$ direction, based on an in-plane AFM ordering of some particular HOMM. On the top surface, we select a sign of rotation such that atomic sites with positive bulk magnetic multipoles now have a $+\hat{z}$ dipole component, and atomic sites with negative magnetic multipoles acquire a $-\hat{z}$ component. Now, note that the surface normal (or effective electric field) of the bottom surface $\mathbf{\hat{n}}_{\mathrm{bottom}}$ points opposite to the surface normal/effective electric field of the top surface ($\mathbf{\hat{n}}_{\mathrm{bottom}}=-\mathbf{\hat{n}}_{\mathrm{top}}$). Thus, if the magnetic multipoles dictating this pattern of canting are odd under inversion, such as the ME multipoles, then induced $+z$ dipole moments on the top surface are symmetrically equivalent to induced $-z$ dipole moments on the bottom surface for a fixed sign of the relevant bulk multipoles corresponding to top surface and bottom surface atoms.  On the other hand, if the HOMMs dictating the AFM out-of-plane canting pattern are even under inversion, such as the magnetic octupoles, a given sign of magnetic octupole will generate the same sign of induced dipole moment on top and bottom surfaces. When we cant in symmetry-disallowed patterns on the other hand, we cant the magnetic dipole moments on the top surface only since there is no clear symmetry connection between the surfaces in this case.\\
\indent For calculations of the bulk atomic-site HOMMs in $\mathrm{CuMnAs}$ and $\mathrm{NiO}$, we use an in-house modified version of VASP which allows us to decompose the density matrix $\rho_{lm,l'm'}$, calculated in DFT, into irreducible (IR) spherical tensor components $w_t^{kpr}$ of the HOMM elements\cite{Spaldin2013}. $k$ refers to the power of the spatial index (e.g., $k=1$ for ME multipoles and $k=2$ for magnetic octupoles); $p$ gives the order of the spin index ($p=0$ for charge, $p=1$ for magnetic multipoles); $r$ is the rank of the IR component of the tensor ($r\in |k-p|,|k-p|+1...k+p$); and $t\in -r,-r+1,...,r$ runs through the tensor components. We then construct the full, \textit{reducible} elements of the ME multipole and magnetic octupole tensors (which are the tensors having a one-to-one correspondence with the magnetoelectric response tensors)  from the IR components based on the forms of reducible tensors given in the next section. The IR components of the atomic-site ME multipoles are defined by $k=1$, $p=1$, and $r=0,1,2$ with the components referred to as the ME monopole, the toroidal moment, and the quadrupole moment respectively. Because ME multipole elements $w_t^{11r}$ switch sign under inversion, only elements in the density matrix with $l+l'$ odd contribute. The IR components of the atomic-site magnetic octupoles with $r=1,2,3$ are called the moment of the toroidal moment density $t_i^{\tau}$, the toroidal quadrupole moment $Q_{ij}^{\tau}$ and the totally symmetric components $O_{lm}$. Magnetic octupoles are symmetric under inversion, and thus only terms in the density matrix for which $l+l'$ is even contribute. As a final detail, for implementation reasons our in-house code only calculates atomic-site multipoles for atomic species which have an on-site Hubbard U within the DFT+U formalism. Thus, for calculations of the atomic site HOMMs of the $\mathrm{Mn}$ atoms in $\mathrm{CuMnAs}$, we include a negligible but nonzero $U$ of $U=0.001\mathrm{eV}$, whereas as mentioned earlier we do not use a Hubbard U correction  for constrained magnetic total energy calculations of $\mathrm{CuMnAs}$. 
\subsection{Form of ME multipole and magnetic octupole tensors}
Here we give the full ME multipole elements $\mathcal{M}_{ij}$ and magnetic octupole $\mathcal{O}_{ijk}$ elements, as discussed and calculated in the main text, in terms of the IR components of the tensors as described in the previous section.  In terms of the rank $0$ ME multipole $a$, the rank $1$ toroidal moment $t$, and the rank $2$ quadrupole $q$, the ME multipole tensor is given by
\begin{equation}
    \mathcal{M}=\begin{pmatrix}
        a+\frac{1}{2}q_{x^2-y^2}-\frac{1}{2}q_{z^2} & t_z+q_{xy} & t_y+q_{xz}\\
        -t_z+q_{xy} & a-\frac{1}{2}q_{x^2-y^2}-\frac{1}{2}q_{z^2} & -t_x+q_{yz}\\
        -t_y+q_{xz} & t_x+q_{yz} & a+q_{z^2}
    \end{pmatrix}.
    \label{eq:MEmult}
\end{equation}
For all 27 components of the reducible octupole tensor $\mathcal{O}_{ijk}$ which is directly proportional to the quadratic ME tensor $\beta$, we refer the reader to Ref.\cite{Urru2022}. Ref.\cite{Urru2022} also explains in detail how to decompose $\mathcal{O}_{ijk}$ into its IR components. For the purposes of this manuscript, here we only give in terms of IR components the six elements $\mathcal{O}_{i=(x,y,z)yy}$ and $\mathcal{O}_{i=(x,y,z)zz}$ which are discussed in the main manuscript in relation to the $(010)$ $\mathrm{CuMnAs}$ and $(1\bar{1}0)$ $\mathrm{NiO}$ surface respectively. In terms of the IR components $t_i^{\tau}$, $Q_{ij}^{\tau}$ and $O_{lm}$ mentioned in the above section, these reducible magnetic octupole elements are given by
\begin{align}
    &\mathcal{O}_{xyy}=-\frac{1}{20}(5O_3+O_1)-\frac{2}{3}Q_{yz}^{\tau}-\frac{2}{3}t_x^{\tau};\\
    &\mathcal{O}_{yyy}=-\frac{1}{20}(5O_{-3}+3O_{-1});\\
    &\mathcal{O}_{zyy}=-\frac{1}{10}(5O_2+O_0)+\frac{2}{3}Q_{xy}^{\tau}-\frac{2}{3}t_z^{\tau};\\
    &\mathcal{O}_{xzz}=\frac{1}{20}(4O_1)+\frac{2}{3}Q_{yz}^{\tau}-\frac{2}{3}t_x^{\tau};\\
    &\mathcal{O}_{yzz}=\frac{1}{20}(4O_1)-\frac{2}{3}Q_{xz}^{\tau}-\frac{2}{3}t_y^{\tau};\\
    &\mathcal{O}_{zzz}=\frac{1}{20}(4O_0).
    \label{eq:octupoles}
\end{align}

\subsection{Proof of symmetry-allowed magnetic octupoles given symmetry-allowed magnetic dipoles}

Here we prove that if a given Cartesian component $i$ of the local atomic magnetic dipole $\mathbf{m}$ is allowed by symmetry, then also the components $\mathcal{O}_{ijj}$, with $j = (x, y, z)$ of the local magnetic octupole tensor are symmetry-allowed. In the following we assume $i = z$, but the same set of arguments holds for $i = x$ and $i = y$ as well. 

The symmetry operations that allow for $m_z$ are the rotations $C_{2z}$, $C_{3z}$, $C_{4z}$, $C_{6z}$, $\mathcal{T} C_{2x}$, $\mathcal{T} C_{2y}$, with $\mathcal{T}$ identifying time-reversal. In principle, also space inversion $\mathcal{I}$ and all the above rotations combined with $\mathcal{I}$ are symmetry operations that allow for $m_z$. However, since both the magnetic dipole and the magnetic octupole are invariant upon inversion, it is sufficient to keep into account only the set of aforementioned rotations. 

By acting with a symmetry operation $\mathcal{R}^{(\tau)}$, where $\tau$ identifies whether the spatial operation is combined with time-reversal ($\tau = 1$) or not ($\tau = 0$), the magnetic octupole tensor transforms as: 
\begin{equation}
    \mathcal{O}'_{i'j'k'} = (-1)^{\tau} \sum_{i j k} R_{i'i} R_{j'j} R_{k'k} \mathcal{O}_{ijk}, 
    \label{eq:oct_trans}
\end{equation}
where $R$ is the matrix representation of $\mathcal{R}$ in Cartesian coordinates. Since $R_{ij} = r_i \delta_{ij}$, with $r_i = \pm 1$, for rotations by $180^{\circ}$ about Cartesian axes (for instance, $R = \text{diag}(-1, -1, 1)$ for $C_{2z}$, thus $r_x = r_y = -1$ and $r_z = +1$), Eq. \eqref{eq:oct_trans} for $C_2$ symmetry operations can be simplified as follows: 
\begin{equation}
    \mathcal{O}'_{ijk} = (-1)^{\tau} r_i r_j r_k \mathcal{O}_{ijk},
    \label{eq:oct_trans_c2}
\end{equation}
implying that the entries do not get mixed with one another.
Using Eq. \eqref{eq:oct_trans_c2} we can compute how each entry of $\mathcal{O}_{ijk}$ transforms under $C_2$ rotations that leave the $z$ component of the magnetic dipole invariant. In Table \ref{tab:oct_trans} we report the value of the pre-factor $(-1)^{\tau} r_i r_j r_k$ in Eq. \eqref{eq:oct_trans_c2} for each entry of $\mathcal{O}_{ijk}$. If $(-1)^{\tau} r_i r_j r_k = +1$, the entry is invariant, thus it is allowed by symmetry, instead if $(-1)^{\tau} r_i r_j r_k = -1$ the entry flips sign when the symmetry operation is applied, hence it is forbidden by symmetry. In Table \ref{tab:oct_trans}, we highlight with red boxes the entries that are simultaneously invariant under all $C_2$ rotations considered here. These entries are $xxz$, $xzx$, $yyz$, $yzy$, $zxx$, $zyy$, and $zzz$.
\begin{table*}[t]
    \centering
    \caption{Value of $(-1)^{\tau} r_i r_j r_k$ for $\mathcal{O}_{ijk}$ obtained for different symmetry operations that leave the $z$ component of the magnetic dipole invariant. Red boxes identify the entries of the octupole tensor that are simultaneously invariant under all symmetry operations listed in the first column.}
    \begin{tabular}{|c|c|c|c|}
    \hline
    & \multicolumn{3}{c|}{Value of $(-1)^{\tau} r_i r_j r_k$ for} \\
    \hline
         Symmetry operation & $\mathcal{O}_{xjk}$ & $\mathcal{O}_{yjk}$ & $\mathcal{O}_{zjk}$ \\ 
         \hline
         & & & \\
         $C_{2z}$ & $\begin{pmatrix} -1 & -1 & \phantom{+}\tikzmark{left}1\tikzmark{right} \DrawBox[thick] \\ -1 & -1 & \phantom{+}1 \\ \phantom{+}\tikzmark{left}1\tikzmark{right} & \phantom{+}1 & -1 \end{pmatrix}$ \DrawBox[thick] & $\begin{pmatrix} -1 & -1 & \phantom{+}1 \\ -1 & -1 & \phantom{+}\tikzmark{left}1\tikzmark{right} \DrawBox[thick] \\ \phantom{+}1 & \phantom{+}\tikzmark{left}1\tikzmark{right} \DrawBox[thick] & -1 \end{pmatrix}$ & $\begin{pmatrix} \phantom{+}\tikzmark{left}1\tikzmark{right} \DrawBox[thick] & \phantom{+}1 & -1 \\ \phantom{+}1 & \phantom{+}\tikzmark{left}1\tikzmark{right} \DrawBox[thick] & -1 \\ -1 & -1 & \phantom{+}\tikzmark{left}1\tikzmark{right} \DrawBox[thick] \end{pmatrix}$ \\
         & & & \\
         \hline
         & & & \\
         $\mathcal{T} C_{2x}$ & $\begin{pmatrix} -1 & \phantom{+}1 & \phantom{+}\tikzmark{left}1\tikzmark{right} \DrawBox[thick] \\ \phantom{+}1 & -1 & -1 \\ \phantom{+}\tikzmark{left}1\tikzmark{right} & -1 & -1 \end{pmatrix}$ \DrawBox[thick] & $\begin{pmatrix} \phantom{+}1 & -1 & -1 \\ -1 & \phantom{+}1 & \phantom{+}\tikzmark{left}1\tikzmark{right} \DrawBox[thick] \\ -1 & \phantom{+}\tikzmark{left}1\tikzmark{right} \DrawBox[thick] & \phantom{+}1 \end{pmatrix}$ & $\begin{pmatrix} \phantom{+}\tikzmark{left}1\tikzmark{right} \DrawBox[thick] & -1 & -1 \\ -1 & \phantom{+}\tikzmark{left}1\tikzmark{right} \DrawBox[thick] & \phantom{+}1 \\ -1 & \phantom{+}1 & \phantom{+}\tikzmark{left}1\tikzmark{right} \DrawBox[thick] \end{pmatrix}$ \\
         & & & \\
         \hline
         & & & \\
         $\mathcal{T} C_{2y}$ & $\begin{pmatrix} \phantom{+}1 & -1 & \phantom{+}\tikzmark{left}1\tikzmark{right} \DrawBox[thick] \\ -1 & \phantom{+}1 & -1 \\ \phantom{+}\tikzmark{left}1\tikzmark{right} & -1 & \phantom{+}1 \end{pmatrix}$ \DrawBox[thick] & $\begin{pmatrix} -1 & \phantom{+}1 & -1 \\ \phantom{+}1 & -1 & \phantom{+}\tikzmark{left}1\tikzmark{right} \DrawBox[thick] \\ -1 & \phantom{+}\tikzmark{left}1\tikzmark{right} \DrawBox[thick] & -1 \end{pmatrix}$ & $\begin{pmatrix} \phantom{+}\tikzmark{left}1\tikzmark{right} \DrawBox[thick] & -1 & \phantom{+}1 \\ -1 & \phantom{+}\tikzmark{left}1\tikzmark{right} \DrawBox[thick] & -1 \\ \phantom{+}1 & -1 & \phantom{+}\tikzmark{left}1\tikzmark{right} \DrawBox[thick] \end{pmatrix}$ \\
         & & & \\
         \hline
    \end{tabular}
    \label{tab:oct_trans}
\end{table*}

Next, we prove that all the highlighted entries in Table \ref{tab:oct_trans} are invariant also under the $C_{3z}$, $C_{4z}$, and $C_{6z}$ rotations listed before. To make the argument general, we take a rotation of angle $\theta$ about the $z$ axis (but neglect the case $\theta = 180^{\circ}$ which we already addressed in Table \ref{tab:oct_trans}): 
\begin{equation}
R = \begin{pmatrix} \cos{\theta} & -\sin{\theta} & 0 \\ \sin{\theta} & \cos{\theta} & 0 \\ 0 & 0 & 1 \end{pmatrix}.
\end{equation}
We focus on the octupole entries $\mathcal{O}_{zjk}$ with $j = k$, which are relevant for the discussion in the main text, but similar arguments can be used for the remaining highlighted entries in Table \ref{tab:oct_trans}, i.e., $xxz$, $xzx$, $yyz$, and $yzy$. First, by applying Eq. \eqref{eq:oct_trans} we note that $\mathcal{O}'_{zzz} = \mathcal{O}_{zzz}$, thus $\mathcal{O}_{zzz}$ is allowed by symmetry. Next, we apply the transformation to the $zxx$ and $zyy$ entries. We have: 
\begin{align}
    \mathcal{O}'_{zxx} &= \cos^2 \theta \, \mathcal{O}_{zxx} + \sin^2 \theta \, \mathcal{O}_{zyy} - 2 \sin{\theta} \cos{\theta} \, \mathcal{O}_{zxy}, \label{eq:trans_zxx} \\
    \mathcal{O}'_{zyy} &= \sin^2 \theta \, \mathcal{O}_{zxx} + \cos^2 \theta \, \mathcal{O}_{zyy} + 2 \sin{\theta} \cos{\theta} \, \mathcal{O}_{zxy}, \label{eq:trans_zyy}
\end{align}
where we used the fact that $\mathcal{O}_{zyx} = \mathcal{O}_{zxy}$. Eqs. \eqref{eq:trans_zxx}-\eqref{eq:trans_zyy} imply that, upon rotations about the $z$ axis, the $zxx$ and $zyy$ components mix with each other and with the $zxy$ entry as well. The latter transforms in this way: 
\begin{equation}
    \mathcal{O}'_{zxy} = \cos{\theta} \sin{\theta} \, \left( \mathcal{O}_{zxx} - \mathcal{O}_{zyy} \right) + \left( \cos^2 \theta - \sin^2 \theta \right) \mathcal{O}_{zxy}.
    \label{eq:trans_zxy}
\end{equation}
To prove that $\mathcal{O}_{zxx}$ and $\mathcal{O}_{zyy}$ are allowed by symmetry we have to prove that (i) $\mathcal{O}_{zxy}$ cannot be allowed by symmetry and that (ii) if $\mathcal{O}_{zxy} = 0$ then $\mathcal{O}_{zxx}$ and $\mathcal{O}_{zyy}$ are necessarily non-zero and invariant under the rotation. To address point (i), let us start by denying the thesis, i.e., by assuming that $\mathcal{O}_{zxy}$ is allowed by symmetry, which means that $\mathcal{O}_{zxy} \ne 0$ and $\mathcal{O}'_{zxy} = \mathcal{O}_{zxy}$. This can be achieved if in Eq. \eqref{eq:trans_zxy} we assume that 
\begin{equation}
    \mathcal{O}_{zxx} -\mathcal{O}_{zyy} = 2 \tan{\theta} \, \mathcal{O}_{zxy}.
    \label{eq:cond}
\end{equation}
By substituting this condition into Eqs. \eqref{eq:trans_zxx}-\eqref{eq:trans_zyy} we obtain: 
\begin{align}
    \mathcal{O}'_{zxx} &= \mathcal{O}_{zyy}, \\
    \mathcal{O}'_{zyy} &= \mathcal{O}_{zxx}.
\end{align}
These relationships can be satisfied in two cases: (a) if both $\mathcal{O}_{zxx}$ and $\mathcal{O}_{zyy}$ are forbidden by symmetry, i.e., if they vanish, or (b) if $\mathcal{O}_{zxx} = \mathcal{O}_{zyy}$. In both cases Eq. \eqref{eq:cond} reads $2 \tan{\theta} \, \mathcal{O}_{zxy} = 0$, which can be satisfied only if $\mathcal{O}_{zxy} = 0$ because we are assuming $\theta \ne 180^{\circ}$. This is in contradiction with the hypothesis that $\mathcal{O}_{zxy}$ is allowed by symmetry, thus it must necessarily be $\mathcal{O}_{zxy} = 0$. Finally, to address point (ii) above, we plug $\mathcal{O}_{zxy} = 0$ into Eqs. \eqref{eq:trans_zxx}-\eqref{eq:trans_zyy} and we get: 
\begin{align}
    \mathcal{O}'_{zxx} &= \cos^2 \theta \, \mathcal{O}_{zxx} + \sin^2 \theta \, \mathcal{O}_{zyy}, \\
    \mathcal{O}'_{zyy} &= \sin^2 \theta \, \mathcal{O}_{zxx} + \cos^2 \theta \, \mathcal{O}_{zyy}.
\end{align}
We emphasize that $\mathcal{O}_{zxx}$ and $\mathcal{O}_{zyy}$ would be forbidden by symmetry if it is not possible to find a condition such that $\mathcal{O}_{zxx}$ and $\mathcal{O}_{zyy}$ are invariant, i.e., $\mathcal{O}'_{zxx} = \mathcal{O}_{zxx}$ and $\mathcal{O}'_{zyy} = \mathcal{O}_{zyy}$. However, in our case such condition exists: indeed $\mathcal{O}'_{zxx} = \mathcal{O}_{zxx}$ and $\mathcal{O}'_{zyy} = \mathcal{O}_{zyy}$ are satisfied simultaneously if $\mathcal{O}_{zxx} = \mathcal{O}_{zyy}$. We conclude that $\mathcal{O}_{zxx}$ and $\mathcal{O}_{zyy}$ are necessarily allowed by symmetry. 

\bibliography{surfaceAFM.bib}
\end{document}